\documentclass[]{elsarticle}
\usepackage[sc,noBBpl]{mathpazo}
\usepackage[mathscr]{eucal}
\usepackage{amsmath,amsfonts,amssymb,amsthm}
\usepackage{graphicx}
\usepackage{mathtools}
\usepackage{color}
\usepackage{tgpagella}
\usepackage{subfigure}
\usepackage{caption}
\usepackage{url}

\theoremstyle{plain}
\newtheorem{theorem}{Theorem}

\newtheorem{definition}{Definition}

\newtheoremstyle{note}{\topsep}{\topsep}{\slshape}{}{\scshape}{}{ }{}
\theoremstyle{note}

\numberwithin{equation}{section}
\numberwithin{theorem}{section}
\numberwithin{lemma}{section}
\numberwithin{proposition}{section}
\numberwithin{corollary}{section}
\numberwithin{remark}{section}

\mathtoolsset{mathic,centercolon}
\newcommand\cG{{\mathcal G}}

\newcommand\scN{{\mathscr N}}

\newcommand\mvector{\boldsymbol}

\newcommand\vp{\mvector{p}}

\newcommand\vv{\mvector{v}}
\newcommand\vw{\mvector{w}}
\newcommand\vx{\mvector{x}}

\newcommand\vA{\mvector{A}}

\newcommand\vX{\mvector{X}}

\newcommand\vvarphi{\mvector{\varphi}}

\newcommand\field{\mathbb}

\newcommand\C{\field{C}}

\newcommand\rmd{\mathrm{d}}

\newcommand\Dt{\frac{\mathrm{d}\phantom{t} }{\mathrm{d}\mspace{1mu}
t}}

\newcommand\Dz{\frac{\mathrm{d}\phantom{z} }{ \mathrm{d}z}}

\newcommand\Dtt{\frac{\mathrm{d}^2\phantom{t} }{\mathrm{d}t^2}}
\newcommand\Dzz{\frac{\mathrm{d}^2\phantom{z} }{\mathrm{d}z^2}}
\newcommand\pder[2]{\dfrac{\partial #1 }{\partial #2}}

\begin{document}
\begin{frontmatter}
\title{Non-integrability of  the Huang--Li nonlinear financial model}

\author{Wojciech Szumi\'nski}
\ead{w.szuminski@if.uz.zgora.pl}
\address{Institute of Physics, University of Zielona G\'ora, Licealna 9, PL-65-407, Zielona G\'ora, Poland}

\begin{abstract}
In this paper we consider Huang--Li nonlinear financial system recently studied in the literature. It has the form of three first order differential equations
\[
\dot x=z+(y-a)x,\quad
\dot y=1-b y-x^2,\quad
\dot z=-x-c z,
\]
where $(a,b,c)$ are real positive parameters. We show that this system is not integrable in the class of functions meromorphic in variables $(x,y,z)$.  We give an analytic proof
of this fact analysing properties the of differential Galois group of variational equations along certain particular solutions of the system. 
\end{abstract}
\date{\small Key words: Huang--Li nonlinear financial system; chaotic finance model; non-integrability; non-Hamiltonian systems;   Poincar\'e sections; differential  Galois theory}

\end{frontmatter}
\section{Introduction}
Application of theory of nonlinear dynamics, especially chaos theory, in economics and financial systems was first suggested by May and Beddington in 1975~\cite{May:1975::,Baumol:1989::}. From that moment researchers  found chaotic behaviour in various existing models and recently  a new   models with very complex dynamic are being created. Examples are the forced van der Pol model~\cite{Chian:2000::,Chian:2005::}, Kaldorian model~\cite{Lorenz:93::,Orlando:16::}, IS-LM model~\cite{Cesare:05::,Fanti:07::}, Goodwin's accelerator model~\cite{Lorenz:02::} and  Huang--Li nonlinear financial model~\cite{Huang:93::}, to cite just a few.  

Recently the last model, mostly refereed as a "chaotic finance system", is intensively studied. It is described by the following three-dimensional system
\begin{equation}
\label{eq:system}
\begin{cases}
\dot x=z+(y-a)x,\\
\dot y=1-b y-x^2,\\
\dot z=-x-c z,
\end{cases}
\end{equation}
where $(x,y,z)$ are time-dependent variables and $(a,b,c)$ are real non-negative parameters. Here $x$ represents the interest rate, $y$ is the investment demand, $z$ is the price index. Parameters $(a,b,c)$ denote saving amount, cost per investment and elasticity demand of commercial markets, respectively.  

The complex behaviour of the  system~\eqref{eq:system}	 has been noted first  by Ma and Chen in 2001~\cite{Ma:2001a::,Ma:2001b::}. Then, variety of the papers were published where the dynamics of this model was investigated by means the various methods and techniques such as: Lyapunov exponents and bifurcation diagrams~\cite{Gao:2009::,Yu:12::};   synchronizations with linear and nonlinear feedbacks~\cite{Zhao:2011::,Zhou:2009::},  adaptive~\cite{Jabbari:2012::}, sliding mode~\cite{Kocamaz:2015::} and passive~\cite{Kocamaz:2015::} control methods~\cite{Kocamaz:2015::};  control via linear, speed and time-delay feedbacks~\cite{Yang:2011::,Chen:2008::}.

To get an idea about the complexity of the system we made several Poincar\'e cross sections. This technique is based on simply intersections of trajectories
with a suitably chosen plane of section. As a result we obtain a pattern on plane
formed from intersection points of phase curves with the intersection plane, that is
easy to visualize and interpret.\begin{figure}[t!]
  \centering  \subfigure[Global portrait]{
    \includegraphics[width=0.48\textwidth]{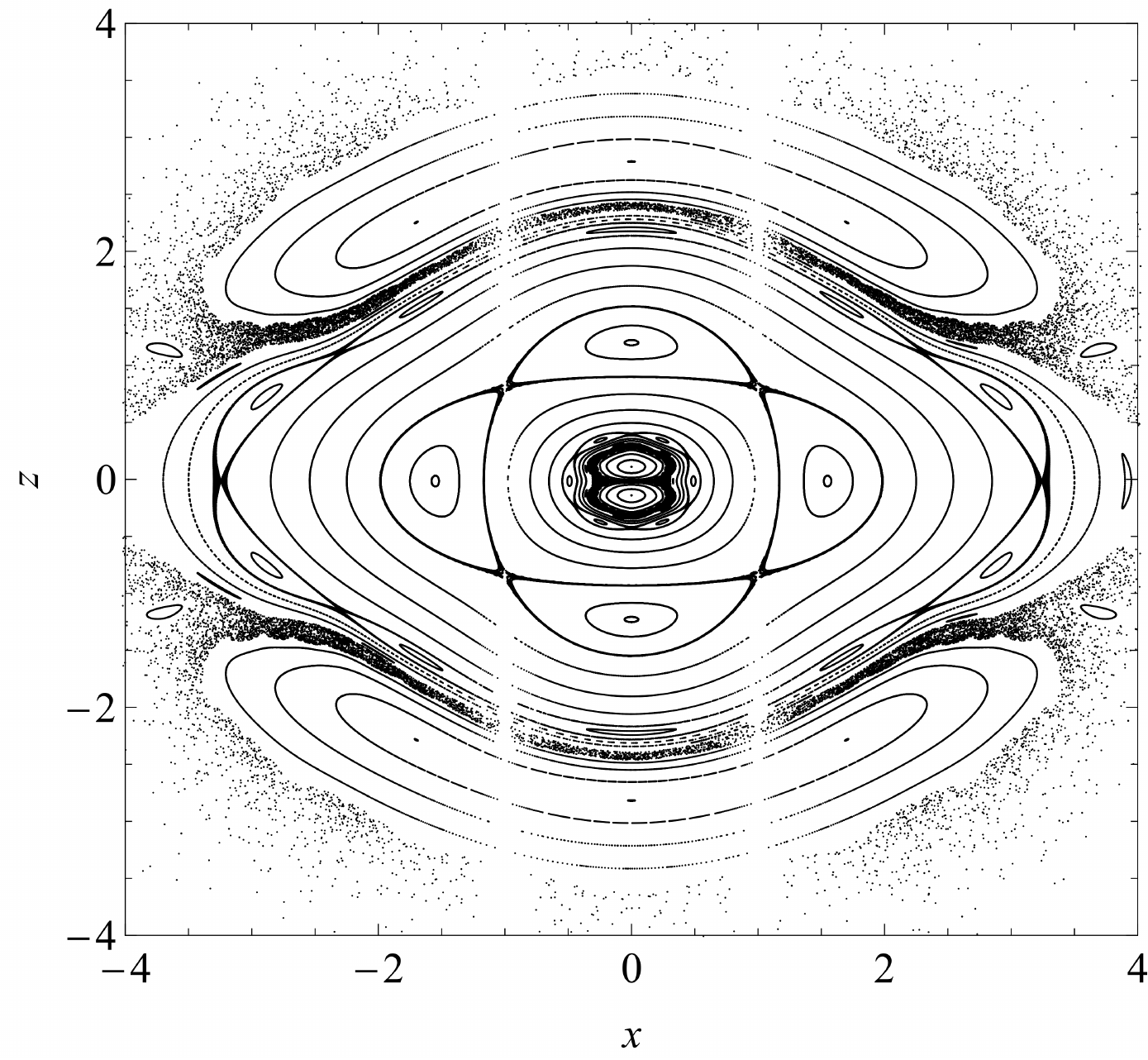}
  }\subfigure[Magnification of  the central region ]{
    \includegraphics[width=0.5\textwidth]{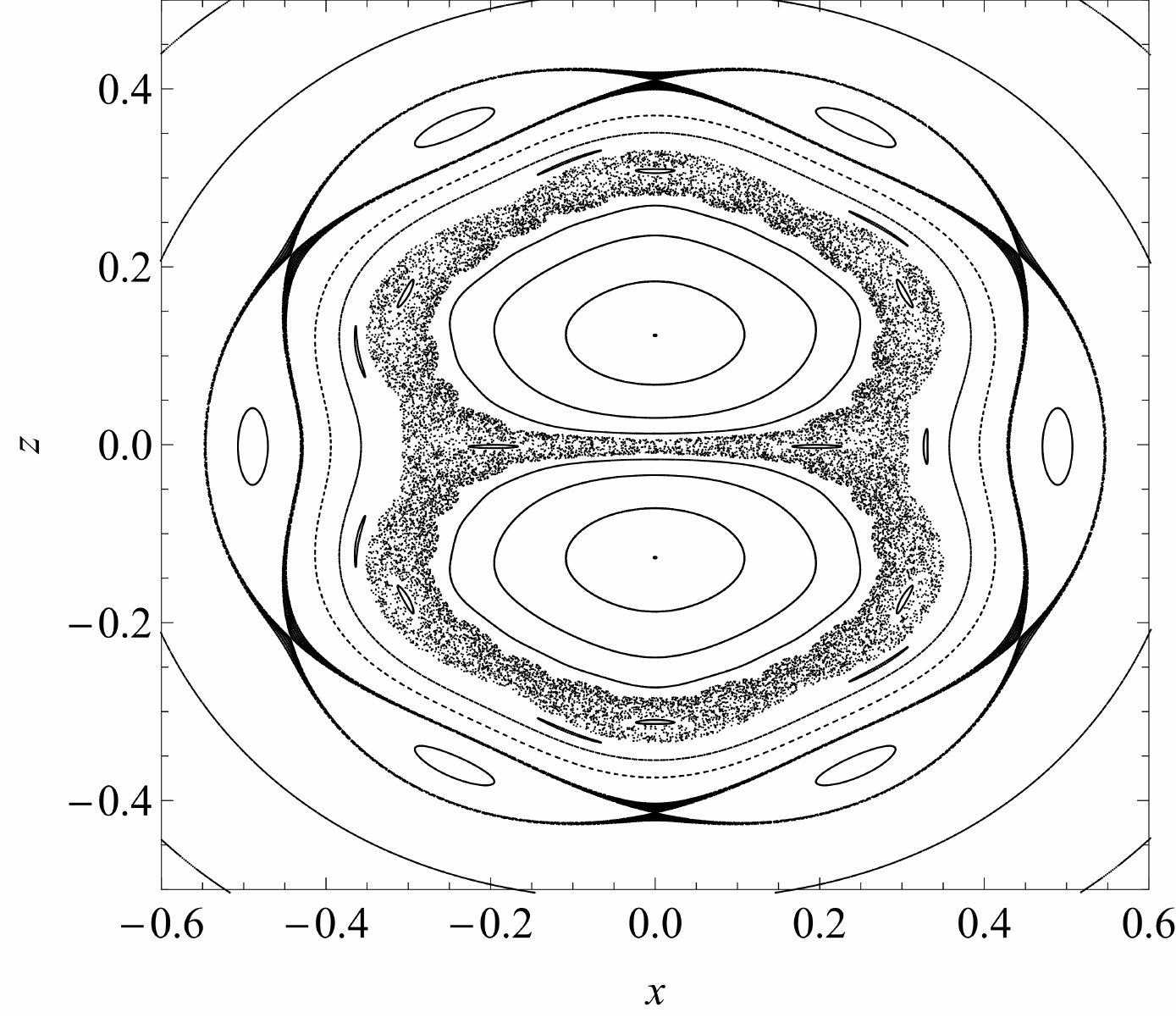}
  }
  \caption{Poincar\'e cross sections of the system~\eqref{eq:system} for $a=b=c=0$\label{fig:1}, on the surface $y=0$}
\end{figure}
\begin{figure}[t!]
\begin{center}
\subfigure[Time series]{
    \includegraphics[width=0.48\textwidth]{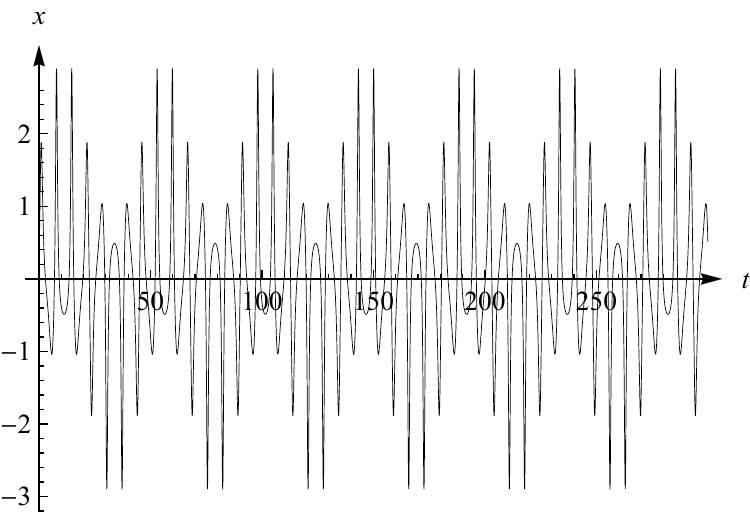}
  }
  \subfigure[Two-dimensional phase portrait]{
    \includegraphics[width=0.48\textwidth]{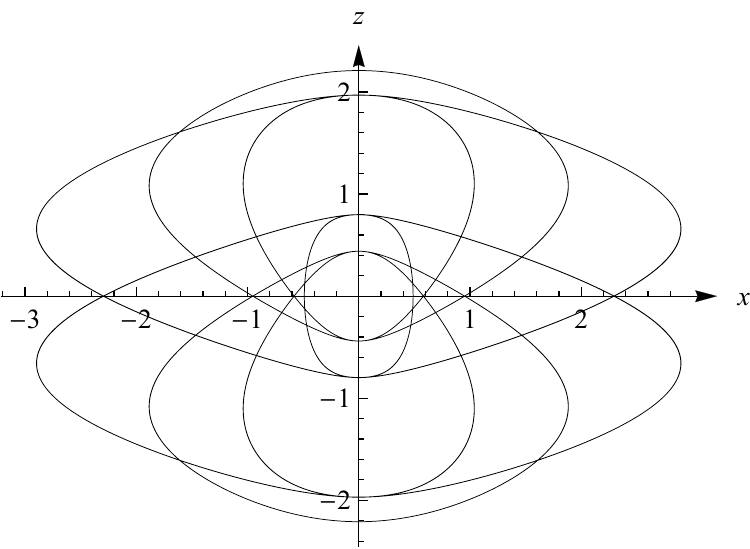}
  }
\caption{Stable periodic motion of the system~\eqref{eq:system} for the parameters:~$a=b=c=0$, with the initial condition: $(x_0,y_0,z_0)=(0,0,2.21)$\label{fig:stable}}
\end{center}
\end{figure}
\begin{figure}[t!]
\begin{center}
\subfigure[Time series]{
    \includegraphics[width=0.48\textwidth]{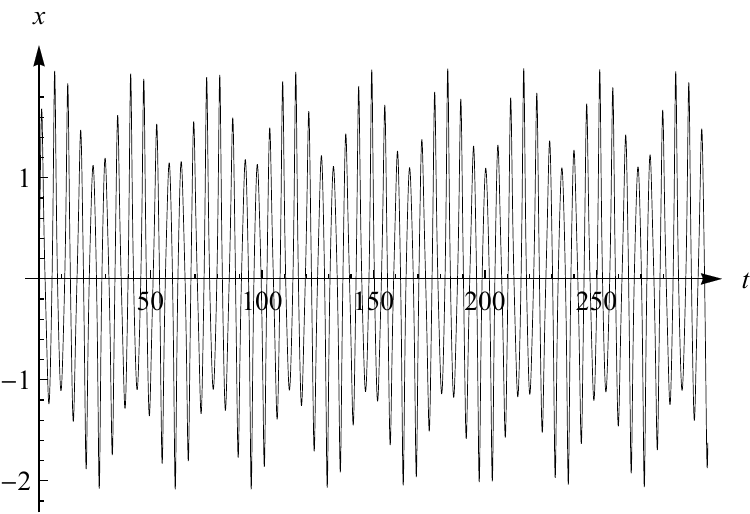}
  }
  \subfigure[Two-dimensional phase portrait]{
    \includegraphics[width=0.48\textwidth]{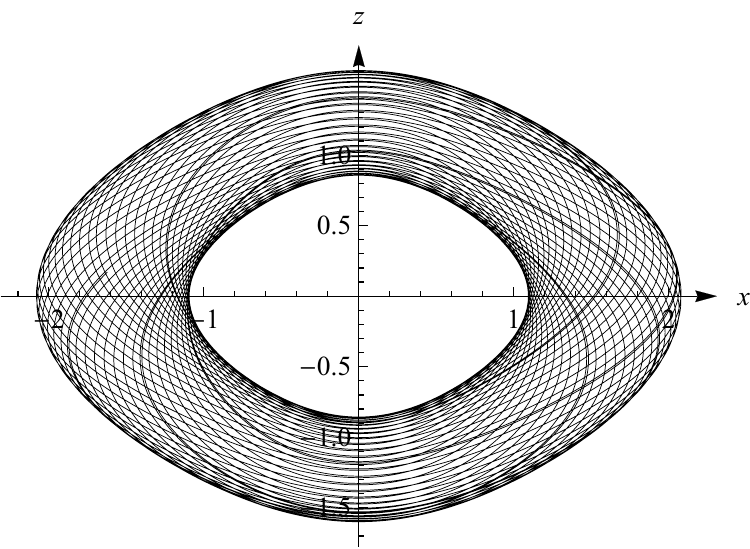}
  }
\caption{Quasi  periodic motion of the system~\eqref{eq:system} for the parameters:~$a=b=c=0$, with the initial condition: $(x_0,y_0,z_0)=(0,0,1.6)$\label{fig:quasi}}
\end{center}
\end{figure}
\begin{figure}[h!]
  \centering \subfigure[Parameters:\ $a=1, b=0.001, c=0.8$ ]{
    \includegraphics[width=0.48\textwidth]{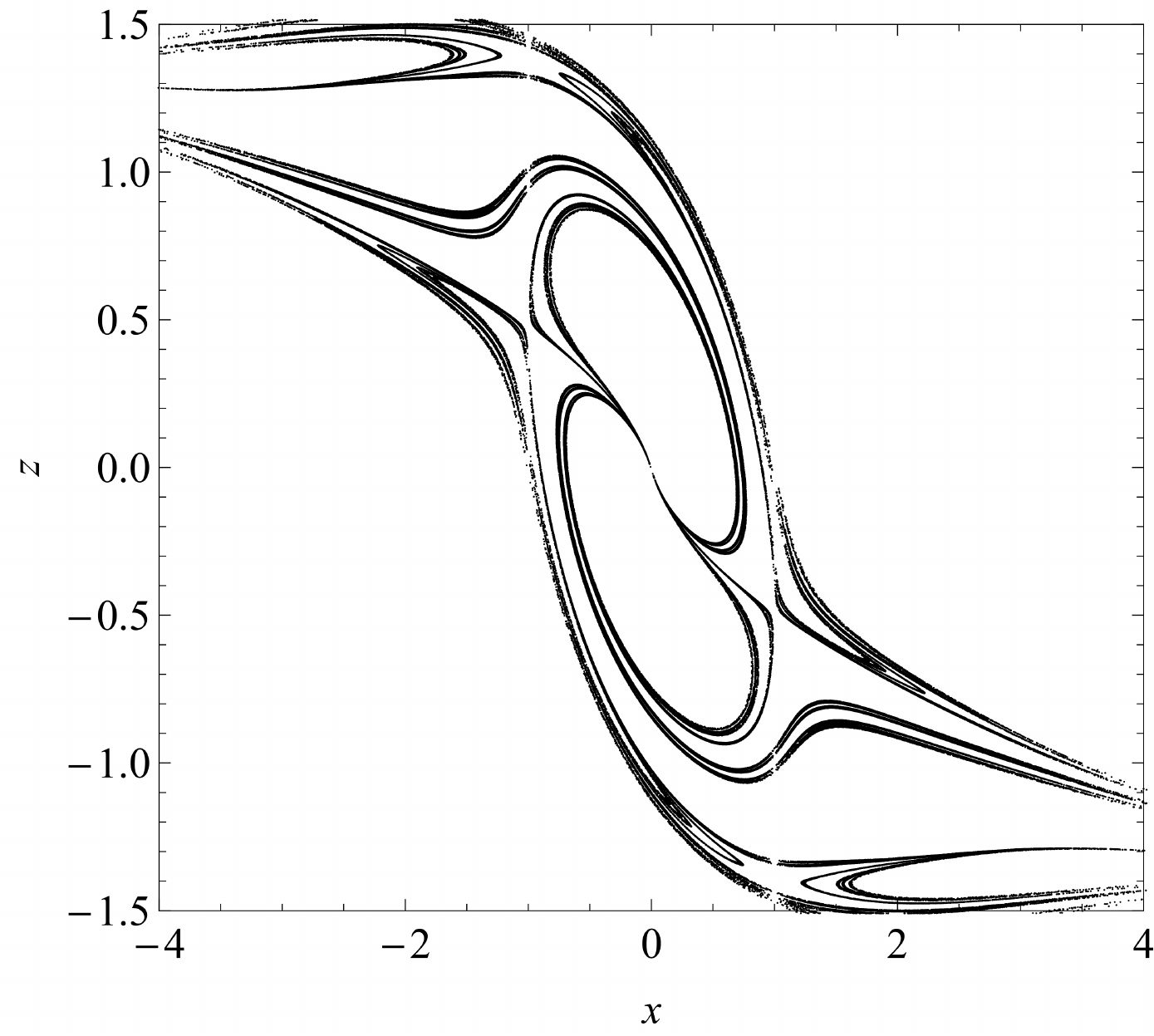}
  }\subfigure[Parameters:\ $a=1, b=0.1, c=0.95$ ]{
    \includegraphics[width=0.48\textwidth]{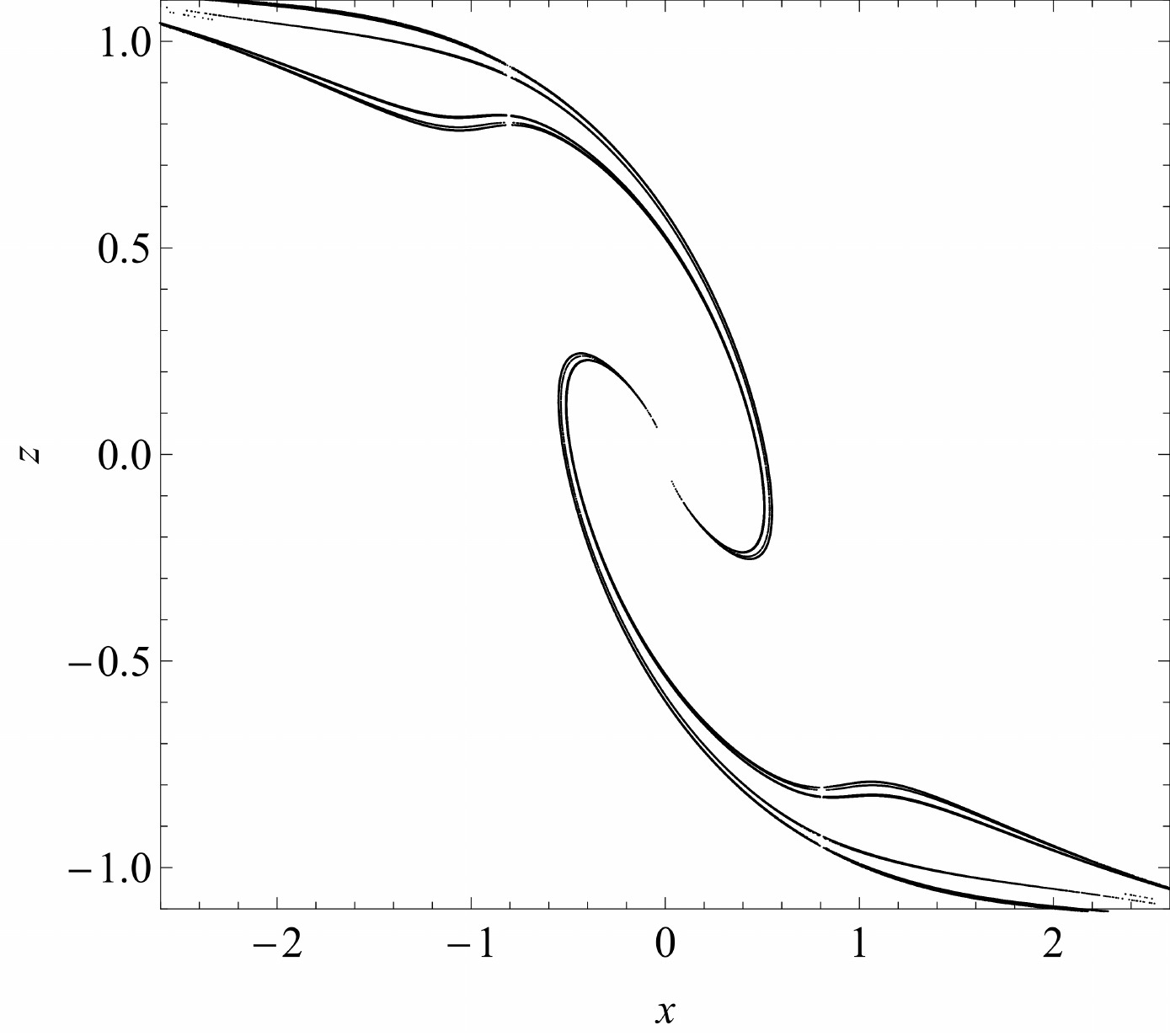}
  } 
  \caption{Poincar\'e cross sections of the system~\eqref{eq:system} on the surface $y=1.5$\label{fig:3}}
\end{figure}
\begin{figure}[h]
\centering
    \includegraphics[width=0.48\textwidth]{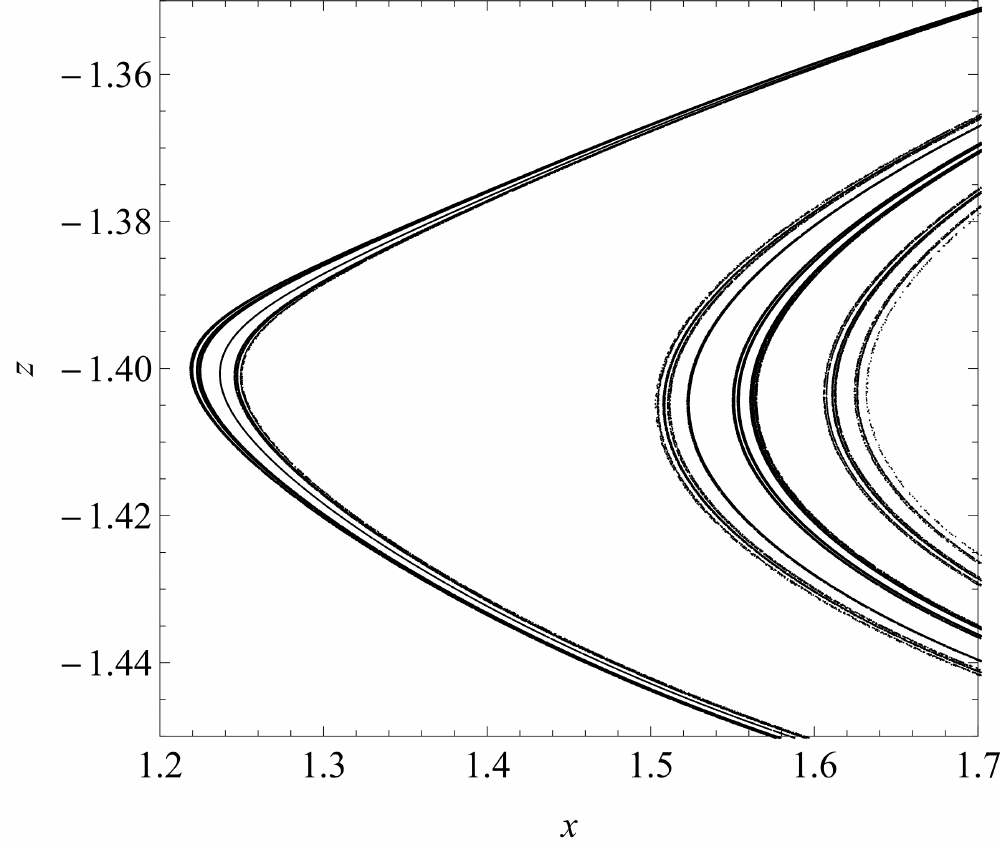}
    \includegraphics[width=0.49\textwidth]{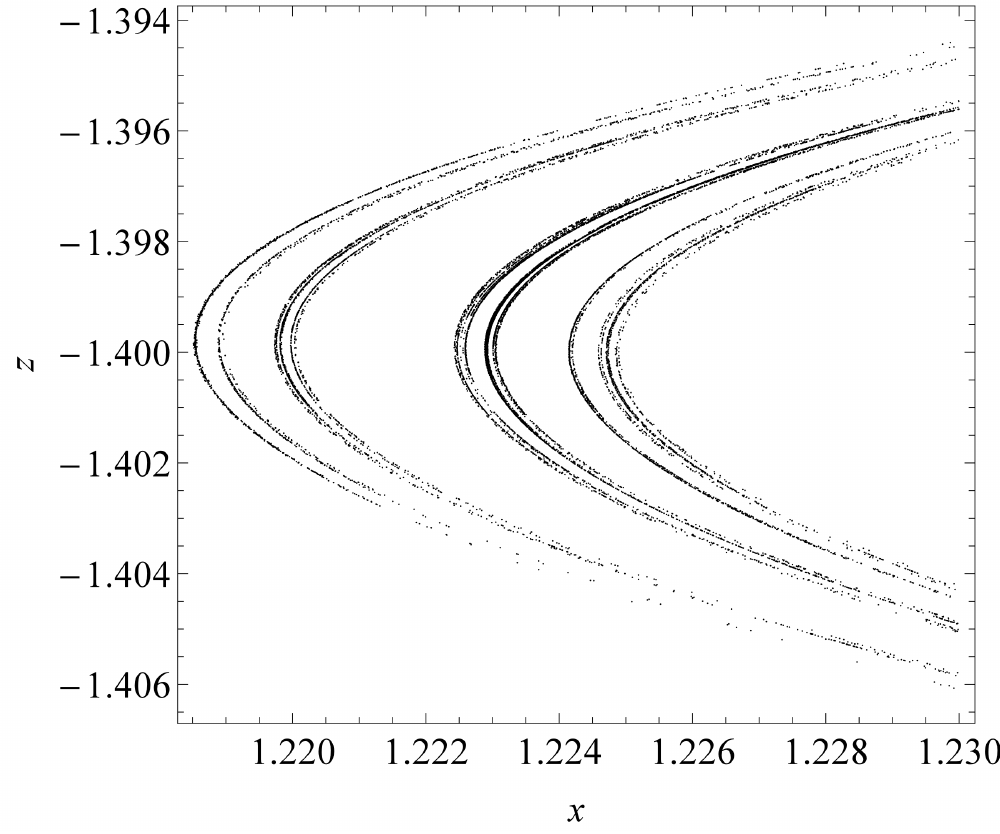}
  \caption{Magnification of  the right bottom region of Fig.~\ref{fig:3}(a) showing  fractals structure.\label{fig:4}}
\end{figure} 
 In Figuers~\ref{fig:1} and~\ref{fig:3} we show such sections. They were generated for certain  values of parameters $(a,b,c)$. The cross-section planes are specified as $y=0$ and $y=1.5$, respectively. The coordinates on these planes are $(x,z)$. Figure~\ref{fig:1} presents first  section and the magnification around its centre. Surprisingly, even for zeroth values of the parameters, the system reveals very rich dynamics. We can detect three types of motion: periodic, quasi-periodic and chaotic, see Figs.~\ref{fig:stable} and~\ref{fig:quasi} presenting periodic and quasi periodic solutions.
In fact the Poincar\'e section visible in Figure~\ref{fig:1} is similar to those for  conservative systems.
 Whereas Figure~\ref{fig:3} posses totally different structure.  For chosen values of parameters  given in captions in Figure~\ref{fig:3}, we obtain  shapely elegant strange attractors of the fractal structure, i.e., their posses the hidden layers structure  preserving self-similarity, see Figure~\ref{fig:4} showing magnifications of the right-bottom region of the Figure~\ref{fig:3}(a).
 
 The complex behaviour of this model apparent from the Poincar\'e cross sections  as well as previously mentioned methods and techniques, suggests its non-integrability. But these numerical signs of non-integrability  were obtained only for chosen values of parameters.
 The high sensitivity of chaotic systems to change the initial conditions makes it impossible to predict the effects of economic decisions in the long time scale.
Therefore, it is crucial  to ask whether there exists any set of values of parameters for which this system is integrable.
 It is very important question from the economical point of view to avoid undesirable trajectories and make the precise economic prediction possible.
 However, it is technically impossible to make the numerical analysis for all values of parameters. For finding  integrable cases
one needs a strong tool to distinguish values of
parameters for which the system is suspected
to be integrable.
 
\section{Integrability analysis}
 For Hamiltonian systems for which we have  a precise notion of integrability, i.e., integrability in the Liouville sense, there are many approaches to the integrability studies:
Hamilton-Jacobi theory, normal forms, perturbation theory, splitting of separateness and recently, the Morales--Ramis theory based on the differential Galois approach. Whereas for non-Hamiltonian systems there is no so many approaches.  It is due to the fact that for non-Hamiltonian systems there is no a commonly accepted definition of the integrability
 Thus, we should first specify what, in the considered context, the integrability means.
\begin{definition}
For a given $n$-dimensional system
\begin{equation}
\label{eq:1}
\dot \vx=\vv(\vx),\qquad \vx=(x_1,\dots, x_n)^T\in \field{C}^n,
\end{equation}
  by B-integrability we understand the existence of $1\leq k\leq n$ functionally independent first integrals
$F_1,\dots, F_k$, and $n-k$ symmetries, i.e., vector fields $\vw_1(\vx)=\vv(\vx),\dots, \vw_{n-k}(\vx)$ such that \begin{equation}
[\vw_i,\vw_j]=0, \quad \text{and}\quad  \vw_j[F_i]=0,\quad \text{for}\quad 1\leq i\leq k,\quad 1\leq j\leq n-k.
\end{equation}
  Then we say that the system is integrable by quadrature.
\end{definition}
In this definition, the second condition means that functions $F_1,\dots, F_k$ are common first integrals of vector fields $\vw_1,\dots , \vw_{n-k}$. It can be shown that if a system~\eqref{eq:1} is B-integrable, then it is integrable by quadrature. That is, all its solutions can be obtained by means of finite sequence of algebraic operations and calculations of primitive functions.

  The  aim of this  letter is to check whether there exist any set of values of  parameters $(a,b,c)$ for which the system~\eqref{eq:system} is integrable. The main result is formulated in
the following theorem.
\begin{theorem}
\label{th:1}
Huang--Li nonlinear financial model~\eqref{eq:system} is not B-integrable  in the class of meromorphic functions of variables $(x,y,z)$ for all real values of  parameters $(a,b,c)$.
\end{theorem}

To prove this theorem we  investigate  variational equations along a particular solution and we study their differential Galois group. This approach of finding necessary conditions for integrability in a framework of differential Galois theory was mostly used in the context of Hamiltonian systems. It is described by Morales--Ramis theory. Thanks to this approach many new integrable cases were detected, see e.g.,~\cite{Pujol:10::,Szuminski:15::,Maciejewski:16::,Maciejewski:16::d}. For a general introduction to differential Galois theory as well as Morales--Ramis theory please consult the papers~\cite{Put:03::,Morales:99::,Maciejewski:09::}.
Although the system~\eqref{eq:system} is not a Hamiltonian, some parts of of the described differential Galois approach to its integrability study can be adopted. 
The key implication is following. If a system~\eqref{eq:system} has  functionally independent meromorphic first integral, then the differential Galois group of variational equations along a particular non-equilibrium solution has a rational invariant. The first applications of the differential Galois theory to non-Hamiltonian systems the interested reader can find in~\cite{Maciejewski:02::,Przybylska:08::}.

 Originally, the Morales--Ramis theory was formulated for Hamiltonian systems for which we identify  integrability as the integrability the Liouville sense.  However, if we restrict ourself to B-integrability, then we have a elegant generalization of this theory. Namely, with the system~\eqref{eq:1} we can also consider its cotangent lift, i.e.,  a Hamiltonian system defined in $\C^{2n}$ with the following Hamiltonian function
\begin{equation}
\label{eq:ham}
H=\sum_{i=1}^np_iv_i(\vx),
\end{equation}
where $\vx=(x_1,\dots, x_n)$ and $\vp=(p_1,\dots, p_2)$ are canonical coordinates defined in a symplectic manifold $M=\C^{2n}$.  In a recent paper~\cite{Ayoul:10::} Ayoul and Zung shown that if the original system~\eqref{eq:1} is integrable, then the lifted system generated by the function~\eqref{eq:ham} is integrable in the Liouville sense.  Hence, for both Hamiltonian and non-Hamiltonian systems, we have the same necessary integrability condition, i.e., the identity component of the differential Galois group of variational equations must be Abelian. We summarize the above facts  by the following theorem that gives necessary integrability conditions for non-Hamiltonian systems.
 \begin{theorem}(Ayoul--Zung)
 \label{th:zung}
 Assume that the system~\eqref{eq:1} is meromorphically B-integrable, then the identity component of the differential Galois group of variational equations along a particular non-equlibrium solution is Abelian. 
 \end{theorem}
\subsection{Proof of Theorem~\ref{th:1}\label{sec:2}}
The system~\eqref{eq:system} possess the invariant manifold
\begin{equation}
\label{eq: invariant manifold}
\scN=\left\{(x,y,z)\in \C^3 | x=z=0\right\}.
\end{equation}
Indeed, equations~\eqref{eq:system} restricted to $\scN$ read
\begin{equation}
\label{eq:particular}
\dot x=0,\quad \dot y=1-b y,\quad \dot z=0.
\end{equation}
Hence solving this equations, we obtain our particular solution $\vvarphi(t)=(0,y(t),0)$.
Let $\vX=[X,Y,Z]^T$ denotes the variations of $\vx=[x,y,z]^T$, then the first order variational equations along $\vvarphi(t)$ take the form
\begin{equation}
\label{eq:variational}
\begin{split}
&\Dt \vX=\vA(t)\vX,\quad \vA(t)=\pder{\vv(\vx)}{\vx}(\vvarphi(t)),
\end{split}
\end{equation}
where the matrix $\vA(t)$ is given by
\[
\vA(t)=\begin{bmatrix}
y-a&0&1\\
0&-b&0\\
-1&0&-c\\
\end{bmatrix}.
\]
Since the particular solution corresponds to a motion along $y$-axis, the equations for $X$ and $Z$ form a subsystem of the normal variational equations that can be rewritten as a one second order differential equation
\begin{equation}
\label{eq:normal}
\ddot X+(a-c-y)\dot X+\left(ac+(b-c)y\right)X=0.
\end{equation}
Next, by means of the change of the independent variable
\begin{equation}
\label{eq:changeindependent}
t\longrightarrow z=\frac{-1+b y(t)}{b^2},
\end{equation}
and using chain formulae for transformation of derivatives
\[
\Dt=\dot z \Dz,\quad \Dtt=\dot z^2\Dzz+\ddot z\Dz,\]
we can rewrite equation~\eqref{eq:normal} as
\begin{equation}
\label{eq:rationalized}
X''+p(z)X'+q(z) X=0,
\end{equation}
where prime denotes derivatives with respect to $z$. The explicit form of the coefficients $p(z)$ and $q(z)$ are the following
\begin{equation}
p(z)=1+\frac{1+b(b-a-c)}{b^2 z},\quad
q(z)=\frac{b-c+abc}{b^3z^2}+\frac{b-c}{bz}.
\end{equation}
The classical change of the dependent variable
\begin{equation}
\label{eq:change dependent}
X=w\exp\left[-\frac{1}{2}\int_{z_0}^zp(s)\rmd s
\right],\end{equation}
 transforms~\eqref{eq:rationalized} into its reduced form
\begin{equation}
\label{eq:reduced}
w''(z)-r(z)w=0,\qquad r(z)=\frac{1}{2}p'(z)+\frac{1}{4}p(z)^2-q(z),
\end{equation}
with coefficient $r(z)$
\begin{equation}
\label{eq:r(z)}
r(z)=\frac{1}{4}-\frac{d+b^2}{2b^2z}+\frac{d^2-b^4-4b^2}{4b^4z^2},
\end{equation}
where \begin{equation}
d:=ab-bc-1.
\end{equation}
In equation~\eqref{eq:reduced} with the coefficient $r(z)$ given above we immediately recognize the Whittaker equation 
\begin{equation}
\label{eq:whittaker}
w''(z)-\left(\frac{1}{4}-\frac{\kappa}{z}+\frac{4\mu^2-1}{4z^2}\right)w(z)=0,
\end{equation}
with
\begin{equation}
\label{eq:parameters}
\kappa:=\frac{d+b^2}{2b^2},\quad \mu:=\frac{\sqrt{d^2-4b^2}}{2b^2}.
\end{equation}
This equation has one regular singularity at $z=0$ and one irregular at $z=\infty$.

Here we should underline one significant fact. Namely, the respective transformations~\eqref{eq:changeindependent} and~\eqref{eq:change dependent} change in general a whole differential Galois group of variational equations~\eqref{eq:variational}. However, the clue is that they do not affect to the identity component of the group, see e.g.,~\cite{Morales:99::}. Thus, according to the  Theorem~\ref{th:zung}, in order to prove a non-integrability of our original nonlinear system~\eqref{eq:system} it is enough to show that the identity component of the differential Galois group $\cG$ of variational equations~\eqref{eq:variational} and thus its rationalized-reduced form~\eqref{eq:reduced} is not Abelian. Necessary conditions for abelianity of the identity component of the
differential Galois group of the Whittaker equation are the following.
\begin{theorem}
\label{th:th2}
The identity component of the Galois group of the Whittaker equation~\eqref{eq:whittaker} is Abelian  if and only if the numbers $(p,q)$ defined by
 \begin{equation}
 \label{eq:kappa and mu}
 p:=\kappa+\mu-\frac{1}{2},\qquad q:=\kappa-\mu-\frac{1}{2}
 \end{equation}
belong to $(\field{N}\times-\field{N}^*)\cup(-\field{N}^*\times\field{N})$.
\end{theorem} 
For details see  Subsection 2.8.3 given in~\cite{Morales:99::}.
According to this theorem
the numbers $(p,q)$ are integer such that one of them is positive and other negative. Thus, its product $pq$ should be always negative. In our case, however, it is easy to verify that  this condition cannot be satisfied. Namely, for $(\kappa,\mu)$ given in~\eqref{eq:parameters}, we obtain the following equality
\begin{equation}
pq=\kappa^2-\kappa-\mu^2+\frac{1}{4}=\frac{1}{b^2},
\end{equation}
which cannot be fulfilled for  $b\geq 0$. This ends the proof.
\section{Conclusions}
Although the obstructions to integrability obtained by means of analysis of  differential Galois group of variational equations are one of the strongest known, the frequent obstacle in its applications is finding a particular non-equilibrium solution for a given dynamical system.  Even though it is a some
limitation, it is weaker than many of the assumptions required in other methods and the result of application of this method is
a non-integrability proof of the considered system finally ending its analysis. Furthermore, if the
system depends on certain parameters (e.g. masses, parameters describing the forces acting on
the system), then usually we can prove its non-integrability for almost all values of parameters
except some finite set of values. In this way the new integrable cases can be found. We shown the application of this approach on the example of intensively studied nonlinear financial model~\eqref{eq:system}. 
We proved that this particular system is not integrable in the class of functions meromorphic in variables $(x,y,z)$ for all values of parameters $(a,b,c)$. 
However, it seems that many others different economical and financial models are waiting for such kind of analysis. It is an open problem.

\section*{Acknowledgement} The author is very grateful to Andrzej J. Maciejewski  and Maria Przybylska for many helpful comments and suggestions concerning improvements and simplifications of some results.
The research has been supported by the grants No. DEC-2013/09/B/ST1/04130 and DEC-2016/21/N/ST1/02477
of National Science Centre of Poland.

\end{document}